\documentclass[12pt]{article}
\pdfoutput=1
\usepackage{putex}

\usepackage{graphicx}
\usepackage{epstopdf}
\usepackage{enumerate}
\usepackage{cite}
\usepackage{tensor}
\usepackage{slashed}
\usepackage{feynmf}
\usepackage{hyperref}
\usepackage{ulem}

\newcommand{\normord}[1]{:\mathrel{#1}:}

\usepackage[usenames,dvipsnames,svgnames,table]{xcolor}

\hypersetup{
	pdfstartview={FitH},    
	pdftitle={Typicality and thermality in 2d CFT},    
	pdfauthor={Datta, Kraus, Michel},     
	colorlinks=true,       
	linkcolor=blue,          
	citecolor=blue!59!black! ,        
	filecolor=magenta,      
	urlcolor=blue!75!black,           
	linktoc=all
}

\numberwithin{equation}{section}

\newcommand {\be} {\begin {equation}}
\newcommand {\ee} {\end {equation}}

\newcommand {\bes} {\begin {equation*}}
\newcommand {\ees} {\end {equation*}}

\newcommand{\beq}{\begin{equation}}
\newcommand{\eeq}{\end{equation}}

\newcommand{\p}{\partial}

\def\be{ \begin{equation} }
\def\ee{ \end{equation} }

\def\qb{\overline{q}}

\def\Tr{\mathop{\rm Tr}}

\newcommand{\bea}{\begin{eqnarray}}
\newcommand{\eea}{\end{eqnarray}}

\def\rt{\rightarrow}

\def\taub{\overline{\tau}}

\newcommand{\ket}[1]{\left| #1\right>}
\newcommand{\bra}[1]{\left< #1\right|}

\begin{document}
	
	
	\institution{UCLA}{ \quad\quad\quad\quad\quad\quad\quad\quad\ \,  Mani L. Bhaumik Institute for Theoretical Physics
		\cr Department of Physics \& Astronomy,\,University of California,\,Los Angeles,\,CA\,90095,\,USA}

	\title{Typicality and thermality in 2d CFT
	}
	
	\authors{Shouvik Datta, Per Kraus and  Ben Michel}
	
	\abstract{We identify typical high energy eigenstates in
		two-dimensional conformal field theories  at finite $c$  and establish that
		correlation functions of the stress tensor in such states are
		accurately thermal as defined by the standard canonical ensemble.
		Typical states of dimension $h$ are shown to be typical level $h/c$
		descendants. In the AdS$_3$/CFT$_2$ correspondence, it is such states that should be compared to black holes in the bulk.  We also discuss the discrepancy
		between thermal correlators and those computed in high energy
		primary states: the latter are reproduced instead by a generalized Gibbs
		ensemble with extreme values chosen for the chemical potentials conjugate to the KdV charges.  }
	
	\date{}
	
	\maketitle
	\setcounter{tocdepth}{2}
	\begingroup
	\hypersetup{linkcolor=black}
	\tableofcontents
	\endgroup
	

	\section{Introduction}
	
	In this paper we address the question of whether typical microstates in two-dimensional conformal field theories appear thermal in a suitable sense.  For a wide range of physical systems, the usefulness of the basic notion of temperature as applied to an isolated system is predicated on this fact, while the eigenstate thermalization hypothesis (ETH) has sharpened the criteria for the emergence of a thermal description \cite{Srednicki:1994,rigol2008thermalization,Garrison:2015lva}.
	
	2d CFTs provide the most tractable class of interacting quantum field theories, so provide a natural arena to address such questions.  On the other hand, this tractability arises due to infinite dimensional Virasoro symmetry, which in turn gives rise to an infinite number of conserved charges that commute with the Hamiltonian (the so-called KdV charges \cite{Bazhanov:1994ft}).   There is an obvious tension between the existence of this infinite tower of charges and the standard description of an ensemble characterized by a finite number of control parameters, i.e. the temperature and chemical potentials.   Hence the notion of a thermal description may need to be refined in this context, for instance by passing to a generalized Gibbs ensemble with an infinite number of chemical potentials.
	
	We focus here on universal aspects of this question.  Namely, suppose
	we are handed a typical energy eigenstate\footnote{States that are not energy eigenstates are also of interest, in particular for studying time evolution towards thermal equilibrium.  We make a few comments on such states in Section \ref{discussion}.}  of the CFT: do correlation
	functions of stress tensors and conserved currents appear thermal in
	such a state, at least in some regime of parameters?  We can answer
	this question without committing to a specific CFT, and if this fails to hold then there will be no effective thermal description of the CFT microstates.
	
	Previous work with similar aims includes \cite{Balasubramanian:2005qu,Balasubramanian:2005mg}.  These papers considered specific theories, namely $\mathcal{N}=4$ super Yang-Mills and the D1-D5 CFT, and considered simplifying features, such as focussing on BPS states or free fields.     Typical microstates were shown to behave approximately ``thermally",\footnote{Thermality here refers not to a physical temperature but to a Boltzmann type factor governing the distribution of states.} with small deviations encoding the specific state.  In bulk language, this  provides evidence that a black hole serves as a coarse grained description of collections of microstates.  As noted above, we proceed here without assuming supersymmetry or making reference to a specific CFT, although we do restrict to two-dimensions and to specific universal probes.  Also relevant is \cite{Balasubramanian:2007qv}, which considers states that are random superpositions of energy eigenstates in a small window, concluding that physically accessible observables have values that are close to thermal, with an error that is exponentially small in the entropy. It was also noted that the nonthermal features can be enhanced to be of order unity by considering imaginary time correlators.  Additional work and reviews include \cite{Alday:2006nd,Balasubramanian:2008da,lai2015entanglement}.

One motivation for this work is to resolve an apparent puzzle regarding a mismatch in the expectation values of KdV charges in microstates versus the thermal ensemble. The simplest example of this mismatch will suffice here.  We consider the stress tensor $T(w)$ along with the conformal normal ordered product $\normord{TT}$, obtained by subtracting power law divergences in the OPE and then taking the coincident limit. The zero modes of these two operators mutually commute, and define the lowest two members of the infinite tower of mutually commuting KdV charges.
 We first consider the CFT on an infinite line at inverse temperature $\beta$, and compute
\bea\label{resa}
\langle T \rangle_{\beta} &=&-{\pi^2 c\over 6\beta^2}~,\cr
 \langle\normord{TT} \rangle_{\beta} &=& \left( {\pi^2 c\over 6\beta^2} \right)^2 +{11\over 90}{\pi^4 c\over \beta^4} ~.
\eea
We next consider the CFT on a spatial circle of circumference $L$.
Let $|h_p\rangle$ denote a Virasoro primary state of dimension $h_p$,\footnote{Here and below we are suppressing dependence on the anti-holomorphic sector of the theory, which for our considerations simply goes along for the ride.} hence obeying $L_0 |h_p\rangle = h_p |h_p\rangle$, $L_{n>0} |h_p\rangle=0$.  The expectation values in this state are
\bea\label{resb}
\langle h_p|T|h_p \rangle &=& -\left(2\pi \over L\right)^2\left(h_p-{c\over 24}\right)~, \cr
 \langle h_p|\normord{TT}| h_p\rangle &=&\left(2\pi \over L\right)^4\left(h_p-{c\over 24}\right)^2 -\left(2\pi \over L\right)^4 \left({h_p\over 6}-{11c \over 1440}\right) ~.
\eea
To compare, we should take $L\rt \infty$ with $h_p/L^2$ fixed so as to maintain a finite energy density in the limit.    Demanding $\langle T \rangle_{\beta}=\langle h_p|T|h_p \rangle$ fixes the relation between  $h_p/L^2$ and $\beta$ as
\bea\label{beta}
 {h_p\over L^2} = {c\over 24\beta^2}~.
\eea
This  gives, in the limit,
\bea
\langle h_p|\normord{TT}| h_p\rangle=  \left( {\pi^2 c\over 6\beta^2} \right)^2~.
\eea
Comparing to (\ref{resa}) we note a discrepancy, which is subleading
at large $c$.  In this work we consider arbitrary $c$, not necessarily
large, in which case the discrepancy is in no sense small.    The same
type of discrepancy persists for quasi-primaries and the higher KdV charges \cite{Basu:2017kzo,He:2017txy}.

One possible response to this discrepancy is that expectation values
computed in the primary state should be compared with those in the
generalized Gibbs ensemble rather than the usual canonical ensemble,
with the infinite number of chemical potentials adjusted to yield
equality for the KdV expectation values.  This avenue has been
explored in \cite{Maloney:2018hdg,Maloney:2018yrz,Dymarsky:2018lhf,Dymarsky:2018iwx,Brehm:2019fyy,Dymarsky:2019etq}.   Here we take  another point of view: we
regard the discrepancy as a reflection of the fact that primary states
are atypical, and we should not expect the canonical ensemble to
accurately reproduce results in such atypical states.  Indeed, in any
system which has a thermal description there will exist atypical
states which appear highly nonthermal.

As we  discuss, a typical state of dimension $h$ is not primary but
rather a typical level ${h\over c}$ descendant of  a dimension $h_p =
{c-1\over c}h$ primary. These states have the form
$|\psi_h\rangle\equiv\prod_n (L_{-n})^{N_n}|h_p\rangle$, $\sum_n
N_n n={h\over c}$, with the $N_n$ being non-negative integers drawn from a Boltzmann distribution, such that $\langle N_n\rangle$ agree with the Bose-Einstein distribution.   We  show that if one chooses a typical state of this form,  then the above discrepancy is resolved: $\langle T\rangle_\beta = \langle \psi_h |T|\psi_h\rangle$ and   $\langle \normord{TT}\rangle_\beta = \langle \psi_h |\normord{TT}|\psi_h\rangle$, where $\beta$ is given by (\ref{beta}) but with $h_p$ replaced by $h$.

We will actually establish a much more general result
\eqref{independence}, namely agreement between stress tensor
correlators computed in the typical microstate versus the thermal
ensemble.  More precisely, consider the case of the two-point function
$\langle T(w)T(0)\rangle$.  For real $w$, corresponding to spatial
and/or real Lorentzian time separation between the two points, the
microstate correlator is accurately thermal provided $L \gg \beta$.
On the other hand, in Euclidean time, corresponding to imaginary $w$,
agreement is present only inside the strip $|{\rm Im}(w)|< \beta$.
For example, the thermal correlator is periodic under $w \rt w
+i\beta$, while this is not even approximately true for the microstate
correlator outside the strip. This is the expected behavior:
 the microstate correlator is singular only in the OPE limit
 $w\rightarrow 0$, while the
 Euclidean periodicity of the thermal correlator implies
 an infinite number of singularities
at $w=in\beta$ for $n\in \mathbb{Z}$. These are ``forbidden singularities'' from the viewpoint of the microstate correlator \cite{Fitzpatrick:2016ive,Faulkner:2017hll}.

The underlying mechanism responsible for the thermal behavior of microstate correlators is the following. A stress tensor correlator involves a weighted sum over expectation values of the form $\langle\psi| L_{n_1} .\ldots L_{n_k}|\psi\rangle$, where $\sum_k n_k =0$. These expectation values vary considerably depending on which microstate we choose. However, the relevant part of the correlator is an infinite sum of the above expectation values multiplied by factors $\cos \left({2\pi n w\over L}\right)$, and what matters is the variance of this object evaluated over the space of dimension $h$ microstates.  We compute this variance in the large $L$ limit and show that it is small provided $\beta$ is held fixed as $L\rt \infty$.    This follows from the fact that $\langle\psi| L_{n_1} .\ldots L_{n_k}|\psi\rangle$ for different choices of the $n_i$  are statistically independent in this regime. Since the variance is small, the correlator takes approximately the same value in a typical microstate as in the thermal ensemble, with corrections suppressed by $1/L$.
As mentioned above, typical  microstate correlators cannot be approximated by thermal correlators outside the strip $|{\rm Im}(w)|<\beta$; this follows from the analytic continuation needed to define the thermal correlators in such cases.

Once we have established that in a typical microstate stress tensor
correlators assume their thermal values, up to small corrections, it
immediately follows that all KdV charges will have nearly thermal
expectation values.  With this in mind we can come back to the
relevance of the generalized Gibbs ensemble.  We have noted that a
typical state of dimension $h$ is based on a primary of dimension
$h_p={c-1\over c}h$, but suppose we instead choose to look at a state
based on a different value of $h_p$.  In this case we do not expect
correlation functions computed in a typical descendant of such a state
to agree with those in the canonical ensemble, but one can ask whether
they would agree with correlators in the generalized Gibbs ensemble
for appropriately chosen potentials.  It would be interesting to
answer this question following the large $c$ analysis in \cite{Dymarsky:2018iwx,Maloney:2018hdg},
but we do not address it here.  The case of a primary state, with
$h_p=h$ is the extreme version of this;  for example,  a primary state
has the lowest value of the second KdV charge among all states of a
given energy. In general, if, for whatever reason, one is interested
in the properties of states (such as primaries) whose KdV charges differ significantly
from their values in the canonical ensemble, then the generalized
Gibbs ensemble is appropriate.  However, since such states are rare,
the introduction of KdV potentials is not necessary to describe most
states, whose expectation values are captured instead by the ordinary
canonical ensemble.

The rest of this paper is organized as follows.  In section \ref{current} we warm up with the technically simple case of spin-1 current correlators.   We present the general argument that typical microstates yield thermal correlators, and then verify this numerically.  In section \ref{stress} we turn to the stress tensor correlators.  The statistical independence of the expectation values of a string of Virasoro generators,  which is the key result needed for approximate thermality, is established in section \ref{statistical} for the case of the two-point function.  The general case is considered in Appendix~\ref{appB}.     We close with some comments in section \ref{discussion}. Appendix~\ref{appA} derives the equivalence between two different forms of the current two-point functions.

\section{Current correlators}
\label{current}

In this section we consider correlation functions of spin-1 currents $J(z)$.  This provides a technically simple context   to compare and contrast correlation functions computed in microstates versus a thermal ensemble.

\subsection{Thermal correlator}

We normalize $J(z)$ such that its 2-point function on the Euclidean plane is
\bea
\langle J(z_1)J(z_2)\rangle = -{1\over (z_1-z_2)^2}~.
\eea
Transforming to the infinite line at inverse temperature $\beta$  via $z = e^{{2\pi \over \beta} w}$ gives
\bea\label{Jbeta}
 \langle J(w_1) J(w_2)\rangle_{\beta} =- {\pi^2/\beta^2 \over  \sinh^2  \left({ \pi (w_1-w_2) \over \beta}\right) }~.
\eea
The current can be realized in terms of a free boson as $J(z)= \p \phi(z)$, where the free boson stress tensor is $T(z) =-{1\over 2} \p \phi\p \phi$. Higher point correlators are obtained from factorization into 2-point functions, as in Wick's theorem.

We next introduce a Euclidean torus with coordinate $w=x+it$ obeying $w\cong w+L \cong w+i\beta$, corresponding to a theory on a spatial circle of size $L$ at temperature $T=1/\beta$.  We write the corresponding torus 2-point function as $\langle J(w_1)J(w_2)\rangle_{L,\beta}$.   $L$ and $\beta $ are interchanged by taking $w\rt iw$, which is the modular S-transformation in terms of the modular parameter
\bea
\tau = {i\beta \over L}~.
\eea
The 2-point function obeys
\bea
\langle J(w_1)J(w_2)\rangle_{L,\beta} = -  \langle J(iw_1)J(iw_2)\rangle_{\beta,L}~.
\eea
The 2-point function is a meromorphic function on the torus with a single pole ${-1\over (w_1-w_2)^2}$.  This, along with the modular property, determines the 2-point function up to a position independent constant.  The constant is determined in terms of the generalized partition function with a chemical potential, $Z(q,y)=\Tr[q^{L_0-{c\over 24}} y^Q]$, where $Q$ denotes the charge corresponding to the current $J$.   This structure arises from Ward identities, and explicit formulas are provided in \cite{Eguchi:1986sb}.  In the case of a free scalar we have
\bea\label{curra}
\langle J(w_1)J(w_2)\rangle_{L,\beta}= -{1\over L^2} \left( \wp(w/L,\tau)+{\pi^2 \over 3}E_2(\tau)-{\pi \over {\rm Im}(\tau) }\right)~,
\eea
where
\bea \wp(w,\tau) = {1\over w^2} + \sum_{(m,n)\neq (0,0)} \left[ {1\over (w+m+n\tau)^2}-{1\over (m+n\tau)^2}\right]
\eea
is the Weierstrass  function
and the Eisenstein series is $E_2(\tau) =1-24 \sum_{n=1}^\infty {nq^n\over 1-q^n}$ with $q=e^{2\pi i\tau}$.  We will use this free boson result in the following, keeping in mind that the general correlator just differs from this by a position independent constant.

For what follows, it will be useful to reexpress the correlator as a mode sum in the free boson theory.   The mode expansion on the cylinder is
\bea
 J(w) = -{2\pi  \over L} \sum_n \alpha_n e^{{2\pi i n w \over L}}~,
 \eea
with
\bea
[\alpha_m,\alpha_n]= m\delta_{m+n,0}~.
\eea
The thermal correlator is
\bea
 \langle J(w_1)J(w_2)\rangle_{L,\beta} ={1\over Z(\tau)}  {\rm Tr} \left[ q^{L_0-{1\over 24}}  \qb^{\tilde{L}_0-{1\over 24}}   J(w_1)J(w_2) \right]~,\quad  q=e^{2\pi i \tau}~,
\eea
with $Z(\tau) =  {\rm Tr} \left[  q^{L_0-{1\over 24}}  \qb^{\tilde{L}_0-{1\over 24}}   \right]$ and
\bea
L_0 = {1\over 2}\alpha_0^2 + \sum_{n=1}^\infty\alpha_{-n}\alpha_n~,\quad
\tilde{L}_0 = {1\over 2}\alpha_0^2 + \sum_{n=1}^\infty\tilde{\alpha}_{-n}\tilde{\alpha}_n~.
\eea
We work in a basis of eigenstates of $\alpha_{-n}\alpha_n$ with
eigenvalues $N_n n$, $N_n$ being the occupation number. In the
canonical ensemble, the probability distribution over occupation numbers is given by the normalized Boltzmann factor,
\bea\label{Pn}
P(N_n) = {e^{2\pi i \tau N_n n}\over \sum_{N_n=0}^\infty e^{2\pi i \tau N_n n} } = (1-e^{2\pi i \tau n})e^{2\pi i \tau N_n n}~.
\eea
The average occupation number is given by the Bose-Einstein distribution
\bea
\langle N_n \rangle_{L,\beta} = \sum_{N_n=0}^{\infty} P(N_n) N_n =  {1\over e^{-2\pi i \tau n}-1}~.
\eea
This yields the thermal correlator\footnote{Here and elsewhere we are implicitly considering the time ordered correlator, with ${\rm Im}(w)>0$.  For ${\rm Im}(w)<0$ the sign of $w$ should be flipped in the formulas below. This ends up being immaterial as the final result is invariant under $w\rt -w$. }
\bea\label{currb}
\hspace{-1cm}\langle J(w)J(0) \rangle_{L,\beta} &=& \left({2\pi \over L}\right)^2 \sum_n \langle \alpha_n \alpha_{-n} \rangle_{L,\beta}  e^{{2\pi i n w  \over L}} ~,\cr
& = & \left({2\pi \over L}\right)^2\left[ \sum_{n>0} n e^{{2\pi i n w \over L} }+\langle\alpha_0^2\rangle_{L,\beta} + 2 \sum_{n>0} \langle \alpha_{-n} \alpha_{n} \rangle_{L,\beta}  \cos \left({2\pi  n w \over L}\right)   \right]  \cr
& = &  \left({2\pi \over L}\right)^2\left[ -{1\over 4 \sin^2 \left({\pi w \over L}\right)}  +{L\over 4\pi \beta}  + 2 \sum_{n>0} {n \over e^{-2\pi i \tau n}-1} \cos \left({2\pi  n w \over L}\right)   \right] ~. 
\eea
Here we have used $\langle\alpha_0^2\rangle_{L,\beta} ={L\over 4\pi \beta}$, as derived in Appendix~\ref{appA}. 

An important point for what follows is that the sum in (\ref{currb}) converges in the strip $|{\rm Im }(w)|<\beta$, due to the competition between the cosine in the numerator and the Bose-Einstein exponential in the denominator, but diverges outside the strip.  Inside the strip the correlator is periodic under $w \rt w+i\beta$, and we use this relation to analytically continue the correlator to the full $w$-plane.

The equivalence of (\ref{curra}) and (\ref{currb}) is shown in Appendix \ref{JJ-forms}.

\subsection{Microstate correlator}

In a microstate, $|\psi\rangle$, the current two-point function takes a
similar form,
\bea\label{currc}
\langle \psi|J(w)J(0)|\psi\rangle =\left({2\pi \over L}\right)^2\left[ -{1\over 4 \sin^2 \left({\pi w \over L}\right)}  +\langle \psi|\alpha_0^2|\psi\rangle  + 2 \sum_{n>0} N_n n \cos \left({2\pi  n w \over L}\right)   \right]~.
\eea
We have assumed that $|\psi\rangle$ is an eigenstate of the number operator, $\alpha_{-n}\alpha_n |\psi\rangle = N_n n |\psi\rangle$ (for $n>0$).

We now ask to what extent the correlator evaluated in a typical
microstate agrees with the thermal correlator at an appropriate
temperature.  First, we need to establish what we mean by a typical
microstate.  As above, we restrict to states that are eigenstates of
$\alpha_{-n}\alpha_n$.  The total energy $E={2\pi \over L}
\sum_{n=1}^\infty N_n n$ is assumed to be large, $EL \gg 1$, and we
define the effective temperature $\beta(E)=\sqrt{\pi L \over 12 E}$
such that ${2\pi \over L} \langle L_0 \rangle_{L,\beta(E)} \approx
E$. Standard statistical reasoning implies that if we choose such a
state at random, the occupation number of the $n$th level will have
the probability distribution $P(N_n)$ as in (\ref{Pn}).  Accordingly, our definition of typicality corresponds to randomly choosing occupation numbers according to this probability distribution.\footnote{The situation here is equivalent to studies of random partitions of large integers and limit shapes of their corresponding Young diagrams \cite{vershik}. The Bose-Einstein distribution also determines the limiting profile of the appropriate Young diagram.} We further impose  $N_n=0$ for sufficiently large $n$, say ${2\pi n \over L}>E$; this is convenient for numerics and also ensures that we consider only states of finite energy.

It is easy to see that the microstate correlator will differ
completely from the thermal correlator outside the strip $|{\rm
  Im}(w)|<\beta$, a point that was emphasized in \cite{Balasubramanian:2007qv}.  To see this, we recall that the mode sum in
(\ref{currb}) diverges outside the strip, and the thermal correlator
is defined there by analytic continuation from inside the strip.  On
the other hand, the microstate correlator is a finite sum since the
total energy is assumed to be finite, and so no issue of nontrivial
analytic continuation arises.   If $N_n \approx \langle
N_n\rangle_{L,\beta}$ then the sum in the microstate correlator looks
approximately like a truncated version of the thermal sum.  While the
two sums can approximately agree inside the strip they will differ
outside it, just as the sum $\sum_{n=1}^N x^n$ for large $N$ will approximately agree with $1/(1-x)$ for $|x|<1$, but looks completely different for $|x|>1$.

With this in mind, we now restrict attention to $|{\rm Im}(w)|<\beta$.
We now argue that the microstate correlator will look approximately
thermal provided $L \gg \beta $.  To see this, we first note that if
we simply insert $N_n = \langle N_n\rangle_{L,\beta}$ along with
$\langle \psi|\alpha_0^2|\psi\rangle={L\over 4\pi \beta}$  in the
microstate correlator, then we reproduce the thermal result.  Of
course, no microstate is precisely compatible with this since $\langle
N_n\rangle_{L,\beta}$ are not integers in general, but we can consider
a microstate for which these relations are approximately true.  Such
microstates are rare, since $N_n$ has large fluctuations over the
space of all microstates of a given energy: by differentiating the partition function $z(q)=\sum_{N_n=0}^\infty q^{N_n n} = (1-q^n)^{-1}$ we have
\bea
{\delta N_n \over N_n} = \sqrt{ \langle (N_n -\langle N_n\rangle)^2 \rangle \over \langle N_n\rangle^2} \approx  q^{-n} = e^{{2\pi \beta n \over L} }.
\eea
This is not small, which implies that $N_n\neq \langle
N_n\rangle_{L,\beta}$ even in typical microstates.

However, the correlator itself an infinite sum of such terms, the relevant piece of which is
\bea
\label{kw}
 K(w) = \left(\frac{2\pi}{L}\right)^2\sum_{n>0} \alpha_{-n}\alpha_n \cos\left(2\pi n w\over L\right)~.
 \eea
We can evaluate the fluctuations in this operator using
\bea
\langle \alpha_{-n}\alpha_n\rangle_{L,\beta} &=& {q^n\over 1-q^n} n, \cr
  \langle \alpha_{-m}\alpha_m\alpha_{-n}\alpha_n  \rangle_{L,\beta} &=&\langle \alpha_{-m}\alpha_m\rangle_{L,\beta}\langle \alpha_{-n}\alpha_n\rangle_{L,\beta} +  {q^{ {2}n}\over (1-q^n)^2}n^2\delta_{m,n} ~.
\eea
 If we take $L\rt \infty$ at fixed $w$ and $\beta$\ \footnote{We relax the condition on $w$ below.}  we can convert sums to integrals and find
\begin{align}
&\left(\frac{L}{2\pi}\right)^2\langle K(w)\rangle_{L,\beta}=    \left(L\over 2\pi \beta\right)^2\int_0^\infty {x\cos \left( {w\over \beta}x\right) \over e^{x}-1} dx =  {L^2 \over 8\pi^2 w^2}-{L^2 \over 8\pi^2 \beta^2}{1\over \sinh\left(\pi w\over\beta\right)^2}~,  \cr
&\left(\frac{L}{2\pi}\right)^4\left[\langle K^2(w)\rangle_{L,\beta} -  \big(\langle K(w)\rangle_{L,\beta} \big)^2\right] ={L^3 \over 8\pi^3 \beta^3} \int_0^\infty{ x^2 \cos^2\left({w\over \beta}x\right) \over (e^x -1)^2} dx~.
\end{align}
We first note that by using the first line and taking $L\rt \infty$ we find that (\ref{currc}) correctly reduces to (\ref{Jbeta}). The integral appearing in the expression for $\langle K^2(w)\rangle_{L,\beta}$ above can be formally evaluated in terms of Hurwitz zeta functions but its explicit form is not illuminating.   We then compute the size of the fluctuations as
\bea
\delta K = \sqrt{ \langle (K - \langle K\rangle)^2 \rangle } = \frac{1}{L^2} \left(L\over \beta\right)^{3/2} f(w/\beta)
\eea
so that $\delta \langle J(w)J(0)\rangle_{L,\beta} \sim {1\over
  \sqrt{L}}$ as $L\rt \infty$, which is just the standard magnitude of
finite size corrections to the thermodynamic limit. Since the sum
(\ref{kw}) is sharply peaked in the ensemble of microstates in this regime,
the correlator in a typical microstate approximates the
thermal result.

The situation changes slightly if we hold fixed $w/L$ as we take $L\rt \infty$.  In this case we cannot replace the sums by integrals due to the relatively rapid variation of the cosines, and we have
\bea
(\delta K(w))^2 = \left(\frac{2\pi}{L}\right)^4{ \sum_{n>0} {n^2 q^n\over (1-q^n)^2}\cos^2 \left(2\pi n w\over L\right)^2}~.
\eea
For $w=0$, or any multiple of $L/2$, the cosine factor becomes unity,
which allows us to replace the sum by an integral, yielding, $\delta
K(w) =\sqrt{2\pi^3 \over 3L\beta^3} $.  On the other hand, if $\Delta
w/L$ is kept nonzero and fixed in the limit, where $\Delta w$ denotes
the distance to the nearest multiple of $L/2$, then the cosine factor
is rapidly varying compared to the rest of the summand, and can be
replaced by its average, namely $1/2$, yielding $\delta K(w)
=\sqrt{\pi^3 \over 3L\beta^3}$.   All that really concerns us is that,
as above, $\delta K \sim \frac{1}{\sqrt{L}}$, and so the fluctuations
  in the correlator are once again suppressed in the large $L$ limit.

These arguments are readily verified by numerical analysis.   To implement this we generate a list of occupation numbers,  $(N_1, N_2, \ldots )$ by drawing numbers according to the probability distribution $P(N_n)$.  We then insert these occupation numbers in the microstate correlator (\ref{currc}) and plot the result.
\begin{figure}[h!]
   \begin{center}
 \includegraphics[width =1\textwidth]{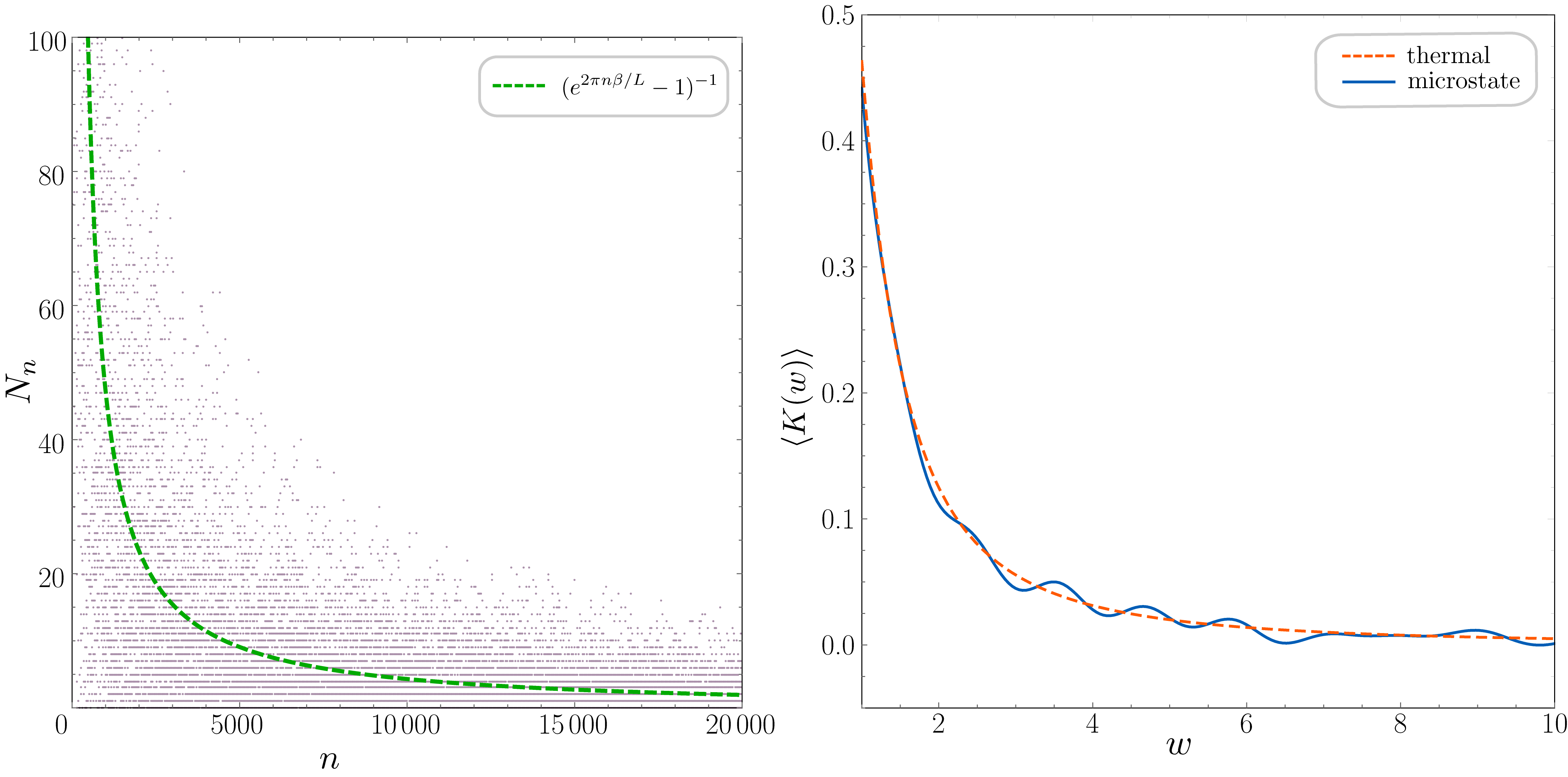}
 \caption{[Left] A random partition, i.e. a set of $N_n$ drawn from the distribution $P(N_n)$ in \eqref{Pn} with  $\beta~=~1, \, L=3\times 10^6$.
  [Right] Comparison of the term that
   differs between the thermal and microstate correlators (the third
   terms in \eqref{currb} and \eqref{currc} respectively). The microstate on the plot
   is defined by the $\{N_n\}$ from the left panel.}
 \label{f2}
 \end{center}
 \end{figure}
\vspace*{-.5cm}

For $|w|\ll L$ the correlators decay exponentially in $|w|$.  However, they must eventually increase to respect the periodicity $w\cong w+L$.  The minimal value is reached for $w \approx L/2$, and as shown in   Appendix \ref{minimal-size},   $\langle J({L\over 2} )J(0)\rangle_{L,\beta} \sim -{\pi \over \beta L} $, which vanishes as $L\rt \infty$ as expected.

It is worth commenting on some related plots that appear in
\cite{Balasubramanian:2005qu} (see their fig.~1). That paper considers the free CFT corresponding to the D1-D5 system at the
symmetric orbifold point.  At large $N$, this theory has a large
degeneracy of Ramond-Ramond ground states, which are chiral primaries.
The coarse grained description of these ground states is dual to the
$M=0$ BTZ black hole, as was verified  by comparison of a two-point
function computed in the two descriptions.  At large $N$ the typical
ground state correlator is well approximated by the coarse grained
correlator for time separation $t < O(\sqrt{N})$.  For larger  $t$ the
correlator displays an erratic behavior that depends sensitively on
the microstate. The common feature in the two examples is the
appearance of a coarse grained description, but the details differ.

\section{Stress tensor correlators}
\label{stress}

We now turn to the case of stress tensor correlators.  The general approach follows  the previous discussion of current correlators, although the details are a bit more involved.  The conclusion is the same: correlators computed in typical microstates look thermal in the appropriate regime of parameters.

\subsection{Two-point functions}

Stress tensor correlators are highly constrained by conformal invariance; in this section we collect a few results.  On the plane we have
\bea
 \langle T(z')T(z)\rangle = {c/2\over (z'-z)^4}~.
\eea
We transform to new coordinates $w(z)$ using
\bea
T(w) = \left( \p_w z\right)^{2}T(z) +{c\over 12} \left( {\p_w^3 z \p_w z \over \p_w z\p_w z}-{3\over 2}\left({\p_w^2 z \over \p_w z}\right)^2 \right)~.
\eea
The correlator on the line at inverse temperature $\beta$ is generated by $z=e^{{2\pi \over \beta}w}$, yielding
\bea\label{TTthermal}
 \langle T(w') T(w)\rangle_\beta =   \left( {\pi^2 c\over 6\beta^2}  \right)^2+{c\over 32}\left({2\pi \over \beta}\right)^4 {1\over \sinh^4(  {\pi \over \beta}(w'-w))}    ~.
\eea
The thermal expectation value of the normal ordered product $\normord{TT}$, obtained by taking the coincident limit $w'\rt w$ after removing singular terms in the Laurent expansion, is
\bea\label{TTnorm} \langle \normord{TT}\rangle_\beta =  \left( {\pi^2 c\over 6\beta^2}  \right)^2+ {11\over 90}{\pi^4 c\over \beta^4} ~.
\eea

The stress tensor two-point function on the torus is fixed by a combination of conformal invariance and knowledge of the torus partition function, the latter quantity depending on the specific CFT.  The two-point function is meromorphic, and so determined up to a constant by its singularities, which are in turn fixed by the OPE, $T(w)T(0) \sim {c/2\over w^4} + {2\over w^2}T(0)+{1\over w}\p T(0)$.  The coefficient of the double pole is therefore fixed by the one-point function, which is in turn given by differentiating the partition function with respect to the modular parameter.  The undetermined constant part of the correlator is fixed by Ward identities.  The explicit formula for the correlator may be found in \cite{Eguchi:1986sb}.    The same logic applies to higher genus Riemann surfaces as well.

Next, we would like the result for the stress tensor two-point function on a spatial circle evaluated in a primary state.  If $O_{h_p}$ is a primary operator then on the plane we have
\bea
  \langle O_{h_p}(0) T(z') T(z) O_{h_p}(\infty)\rangle
 = {h_p^2\over z^2 z'^2} +{2h_p\over zz'(z'-z)^2}+{ c/2\over (z'-z)^4}~.
\eea
This is fixed by the conformal Ward identity for stress tensor insertions (or, equivalently, by the fact that it must be a meromorphic function of $z$ and $z'$, with singularities fixed by the OPE). As usual, $O_{h_p}(\infty) = \lim_{z\rt \infty} z^{2h_p} O_{h_p}(z)$.  We now transform to the cylinder with a spatial circle of circumference $L$ via $z=e^{{2\pi i \over L}w}$, which gives
\bea\label{TTmicro}  \langle h_p| T(w) T(0) |h_p\rangle_L = \left(2\pi \over L \right)^4 \left(h_p-{c\over 24}\right)^2    -\left({2\pi\over L}\right)^4 \left({h_p\over 2 \sin^2 ({\pi w\over L} ) } -{c\over 32 \sin^4 ({\pi w \over L} ) } \right)~.
\eea

A naive test of thermality consists of comparing (\ref{TTthermal}) to
(\ref{TTmicro}) in the thermodynamic limit.  In particular we take
$L\rt \infty$ while simultaneously holding $h_p/L^2$ fixed to maintain a finite energy density.   For the correlators to match at large separation, which yields $\langle T\rangle^2$, we should take
\bea {h_p\over L^2} = {c\over 24\beta^2}
\eea
in the limit. The primary state result becomes
\bea
 \langle h_p| T(w) T(0) |h_p\rangle_{L\rt \infty} = \left( {\pi^2 c\over 6\beta^2}  \right)^2-{\pi^2 c\over 3\beta^2}  {1 \over w^2 } +{c/2\over w^4}~.
 \eea
Comparing to (\ref{TTthermal}) we see that the two results share the same short distance singularities  and (by construction) long distance limit, but differ otherwise.  For example, the primary state result yields
\bea
\langle h_p| \normord{TT}|h_p\rangle_{L\rt \infty} =  \left( {\pi^2 c\over 6\beta^2}  \right)^2~,
\eea
as opposed to (\ref{TTnorm}).  On general grounds, we expect that in the thermodynamic limit expectation values computed in typical states should agree with those computed in the thermal ensemble, and so the mismatch is an indication that primary states are not typical.  On the other hand, the mismatch goes away at large $c$, indicating that in this regime primary states are typical.

\subsection{Typical states}

The Hilbert space of a two-dimensional CFT can be decomposed into
representations of the Virasoro algebra.  Each conformal family is labelled by a primary operator of some conformal dimension $h_p$ and consists of the primary state $|h_p\rangle$ and its conformal descendants obtained by acting with strings of $L_{-n}$ operators. We consider unitary representations at $c>1$ with no null states.  The full CFT has both left and right moving Virasoro algebras, but since we will only be considering correlators of $T(z)$ we can restrict attention to one chiral half.

To characterize the typical state $\ket{\psi_h}$ at some specific $h\gg 1$,
we note that there are two competing effects. On the one hand, the number of primary states grows exponentially with $h_p$, but on the other hand so too does the number of descendant states at level $h-h_p$.  The typical value of $h_p$ will be the one that balances these effects.

We write the partition function of the CFT as 
\bea
Z(q) = {\rm Tr}[ q^{L_0 -{c\over 24}} ] = \sum_{h_p} d(h_p) {\Tr}_{h_p} [ q^{L_0 -{c\over 24}} ]~.
\eea
 The anti-holomorphic dependence  is not made explicit; in what follows, the correlation functions of the stress tensor and/or its modes will be determined by holomorphic derivatives ($\partial_\tau$ or $\partial_q$) of the partition function. $\Tr_{h_p}$ above denotes a trace over states in the conformal family labelled by the primary of weight $h_p$, and $d(h_p)$ is the number of primaries at weight $h_p$.  The corresponding Virasoro character $Z_{h_p}$  is
\bea
Z_{h_p}(q) = {\Tr}_{h_p} [ q^{L_0 -{c\over 24}} ]= {q^{h_p-{c-1\over 24}}\over \eta(q)}  ~,
\eea
where $\eta(q) = q^{1/24} \prod_{n=1}^\infty (1-q^n)$ is the standard Dedekind eta function.  Writing $q=e^{-{2\pi \beta \over L}}$, at high temperature we have
\bea\label{Zp}
\ln Z_{h_p}(q) \approx -{2\pi \beta \over L} \left(h_p-{c-1\over 24}\right)+{\pi L\over 12\beta}~,\quad (\beta \rt 0)
\eea
as follows from the modular behavior of the eta function.   The high temperature behavior of the full partition function is obtained by modular transformation of the vacuum contribution,
\bea Z(q) \approx e^{ \pi c L \over 12 \beta}
~,\quad (\beta \rt 0)~.
\eea
These imply the asymptotic degeneracy of primaries \cite{Kraus:2016nwo}
\bea
d(h_p) \approx e^{2\pi \sqrt{ {c-1\over 6}h_p}}~,
\eea
which takes the same form as the Cardy density of states \cite{CARDY1986186}, except with the replacement $c\rt c-1$.

Next, for a given primary state of weight $h_p$, we need to count up the number of descendant states at level $h-h_p$.  This corresponds to
the number of partitions of $h-h_p$, which is given by the Hardy-Ramanujan formula, $e^{2\pi \sqrt{ {h-h_p\over 6} }}$.   Altogether, the number of states which are level $h-h_p$ descendants of weight $h_p$ primaries are
\bea
d(h;h_p) \approx e^{2\pi \sqrt{ {c-1\over 6}h_p}+2\pi\sqrt{ {h-h_p\over 6} } }~.
\eea
Maximizing with respect to $h_p$ gives
\bea
h_p = {c-1\over c}h~.
\eea
At large $c$, the typical state is nearly primary in the sense that
$h_p \approx h$.  However, we will not be making any such large $c$
assumption here. At finite $c$, the typical states with weight $h$ are
level $h/c$ descendants of a weight $h_p$ primary.\footnote{We still
  need to specify how the descendant level is partitioned; as in
  section~\ref{current}, not all partitions are typical. Thermal and
  microstate correlators in descendent states have recently been
  compared in \cite{Guo:2018pvi}, but the partitions
  of the descendent level considered there are atypical according to
  the notion of typicality that we use.}

\subsection{Typical state two-point function}

On the Euclidean cylinder with a spatial  circle, $w\cong w+L$, the mode expansion of the stress tensor is
\bea
T(w) = -\left({2\pi \over L}\right)^2 \left(L_0-{c\over 24}\right) - \left({2\pi \over L}\right)^2 \sum_{n\neq 0} L_n e^{2\pi i n w\over L}~,
\eea
where the generators obey the Virasoro algebra
\bea
[L_m,L_n]= (m-n)L_{m+n} +{c\over 12}(m^3-m)\delta_{m+n,0}~.
\eea
Let $|\psi_h\rangle$ be an eigenstate of $L_0$,  $L_0|\psi_h\rangle = h|\psi_h\rangle$.  Using the mode expansion and the commutation relations it is straightforward to derive the following expression for the two-point function in such a state,
\bea\label{TTstate}
 \langle \psi_h|T(w) T(0)|\psi_h\rangle  & =&\left(2\pi \over L \right)^4 \left(h-{c\over 24}\right)^2    -\left({2\pi\over L}\right)^4 \left({h\over 2 \sin^2 ({\pi w\over L} ) } -{c\over 32 \sin^4 ({\pi w \over L} ) } \right)  \cr &&\quad  +2  \left(2\pi \over L\right)^4 \sum_{n>0} \langle \psi_h|L_{-n} L_{n} |\psi_h\rangle \cos\left({2\pi n w\over L}\right)~.
\eea
For example, suppose that $|\psi_h\rangle$ is primary, so that $L_{n>0}|\psi_h\rangle=0$ and the second line vanishes.  We then recover (\ref{TTmicro}).

We wish to evaluate this for a typical state.  As discussed in the
previous section, a typical state with weight $h$ is a level ${h\over
  c}$ Virasoro descendant of a primary state $|h_p\rangle$ whose
dimension $h_p = {c-1\over c}h$.  The expectation value of $L_{-n}L_n$
depends on which particular descendant state we choose.  However, we
will show in section~\ref{statistical} that in the thermodynamic limit
the variance of \eqref{TTstate} over the ensemble of such states is small.
Therefore, the expectation value in such states can be approximated by
an average weighted by a Boltzmann factor, with the temperature chosen
so as to yield the desired average weight.   Let $\langle X\rangle_{h_p,\beta}$ denote the average of $X$ defined in this sense, %
\bea
\langle X\rangle_{h_p,\beta} = {1\over Z_{h_p}(q)} {\Tr}_{h_p}[ q^{L_0-{c\over 24}} X ]~.
\eea
Here $q=e^{-{2\pi \beta \over L}}$ as before.    $\beta$ is fixed by demanding $\langle L_0\rangle_{h_p,\beta}= h$, which can be written as
\bea
-{L\over 2\pi} \p_\beta \ln Z_{h_p}(q) = h~.
\eea
Using (\ref{Zp}), valid in the relevant thermodyamic limit, along with $h_p = {c-1\over c}h$, we find
\bea
\beta = \sqrt{c\over 24h}L~.
\eea
Next, we need $\langle L_{-n}L_n \rangle_{h_p,\beta}$ in the thermodynamic limit.  As derived in the next section, the result is
\bea\label{LL}
\langle L_{-n}L_n \rangle_{h_p,\beta} &=& {1\over e^{{2\pi \beta n \over L}}-1} \left[{c\over 12}n^3+ \left(h_p+{L^2 \over 24\beta^2}\right)2n\right] \cr
&=& {1\over e^{{2\pi \beta n \over L}}-1} \left[{c\over 12}n^3+2hn\right]~.
\eea
As argued above, provided $L \gg \beta$, in a typical state we can make the following replacement in (\ref{TTstate}):
\bea\label{repl}
 2  \left(2\pi \over L\right)^4\sum_{n>0} \langle \psi_h|L_{-n} L_{n} |\psi_h\rangle \cos\left({2\pi n w\over L}\right) & \rt &  2  \left(2\pi \over L\right)^4\sum_{n>0}\langle L_{-n}L_n \rangle_{h_p,\beta} \cos\left({2\pi n w\over L}\right)~. \cr
 &&
 \eea
Using (\ref{LL}), converting the sum to an integral at large $L$, and using
\bea
\int_0^\infty {x^3+ 4\pi^2 x\over e^x-1} \cos(ax) dx = -{3\over a^4}+{2\pi^2 \over a^2}+{3\pi^4\over \sinh^4(\pi a)}
\eea
we find
\bea
\langle \psi_h| T(w)T(0)|\psi_h\rangle =\left( {\pi^2 c\over 6\beta^2}  \right)^2 + \left({\pi^4 c\over 2\beta^4}\right) {1\over \sinh^4(  {\pi w\over \beta}) }~.
\eea
This reproduces the thermal correlator in (\ref{TTthermal}), thus verifying that the stress two-point function in a typical state appears thermal, provided $L\gg  \beta$.  It immediately follows that $\langle \psi_h |\normord{TT}|\psi_h\rangle = \langle \normord{TT}\rangle_\beta$.

The key step in obtaining this result was the replacement  (\ref{repl}), whose validity depends on the microstate expectation value being sharply peaked over the ensemble of states.  Obtaining analogous results for higher point correlators of the stress tensor will similarly depend on establishing that operators built out of sums of more $L_n$ are similarly sharply peaked.  We turn to these questions in the next section.

\section{Statistics of Virasoro generator expectation values}
\label{statistical}

We shall now consider the following quantity
\be
X(w) \equiv \sum_{n>0}
X_n \cos\left(\frac{2\pi n w}{L}\right) \equiv \left(2\pi\over L\right)^4
\sum_{n>0} L_{-n} L_n \cos\left(\frac{2\pi n w}{L}\right) .
\ee
The replacement (\ref{repl}) is valid if $X(w)$
is sharply peaked over the thermal ensemble. In order to verify this,  we will
study its fluctuations
\bea
\label{fw}
\delta X^2= \langle X^2\rangle - \langle X\rangle ^2.
\eea
Given the form of the two-point function in (\ref{TTstate}),
we can make the replacement (\ref{repl}) in typical states provided
$ \delta X  \rt  0$ as $L \rt \infty$.  In this section averages are computed by summing over states in a single conformal family, $\langle \ldots \rangle = {1\over Z_{h_p}} {\rm Tr}_{h_p} [q^{L_0-{c\over 24}} \ldots ]$, although all formulas are unchanged if there happen to be multiple primaries of weight $h_p$.

The contributions to (\ref{fw}) can be split into off-diagonal and diagonal pieces
\be
\delta X^2= \delta X^2_\text{off-diag}(w) + \delta X^2_\text{diag}(w)
\ee
where
\be
\label{offdiag}
\delta X^2_\text{off-diag} = \sum_{m\neq n} \left(\langle X_m X_n\rangle - \langle
X_m\rangle \langle X_n\rangle\right) \cos\left(\frac{2\pi m
	w}{L}\right) \cos\left(\frac{2\pi n w}{L}\right)
\ee
and
\be
\delta X^2_\text{diag}=\sum_{n>0} \left(\langle X_n^2\rangle - \langle X_n\rangle^2\right) \cos^2\left(\frac{2\pi n
	w}{L}\right).
\ee
The mode number $n$ is taken to be of order $n\sim L/\beta$.  The relevance of this scaling follows from the fact that when we convert sums to integrals we write  $x=\frac{2\pi \beta n}{L}$.  The Bose-Einstein factor then appears as $(e^x-1)^{-1}$,  leading to exponential suppression of the $x\gg 1$ regime.  The same was true in the current correlator case.

Let us first consider (\ref{offdiag}). To evaluate such expectation values we will repeatedly make use of a
simple trick: using
\bea
L_n q^{L_0} = q^{L_0+n} L_n
\eea
and cyclicity of the
trace, one can move the leftmost operator to the right end of the
string and then rewrite the resulting expression as a sum of commutators plus the
expectation value of the original string. For example, to compute
$\langle X_n\rangle$ we write
\bea
 \langle L_{-n} L_n \rangle = q^n \langle L_n L_{-n}\rangle
= {q^n}\left(\langle L_{-n} L_n\rangle + \langle [L_n, L_{-n}]\rangle \right)~.
\eea
Then
\bea
\label{nn}
\langle X_n \rangle  &=& \left(2\pi \over L\right)^4\frac{q^n}{1-q^n}  \langle [L_n, L_{-n}]\rangle \cr
&=& \left(2\pi \over L\right)^4\frac{q^n}{1-q^n} \left(2nq \p_q \ln Z_{h_p} + \frac{c}{12}(n^3-n)\right).
\eea
This is an exact formula.  Now, for large $L/\beta$ we have $\ln Z_{h_p}  \sim L/\beta$ so $q\p_q \ln Z_{h_p}\sim \left({L\over \beta}\right)^2$,    and from this we find the leading behavior $\langle L_{-n}L_n\rangle \sim  \left({L\over \beta}\right)^3$, where we have included the $n\sim L/\beta$ scaling.

Similarly, to compute $\langle X_n X_m\rangle$ we write
\bea
\label{nnmm_rearrange}
\langle L_{-n} L_n  L_{-m} L_m\rangle&=& \frac{q^n}{1-q^n}
\left(\langle  L_n
L_{-m}[L_m,L_{-n}] \rangle + \langle L_n [L_{-m}, L_{-n}]
L_m\rangle \right.\cr & \quad &\quad \quad\quad+ \left. \langle [L_n, L_{-n}] L_{-m} L_m\rangle\right).
\eea
We now show that for $m\neq n$ the first two terms are subleading compared
to the third in the thermodynamic limit.  After evaluating the commutators, each of the three terms is proportional to an expectation value of the form $\langle L_m L_n L_p\rangle$ with $m+n+p=0$.  In the third term one of $(m,n,p)$ equals $0$, unlike for the first two terms.  If none of $(m,n,p)$ equals $0$ then we compute
\bea
\langle L_m L_n L_p\rangle = {1\over 1-q^p} \left[ (p-n)\langle L_m L_{-m}\rangle + (p-m)\langle L_{-n}L_n\rangle \right]~.
\eea
This implies the leading behavior $\langle L_m L_n L_p\rangle \sim \left({L\over \beta}\right)^4$ for this case.
On the other hand if $m=0$ (say), then we have
\bea
\langle L_0 L_{-p}L_{p}\rangle =  \left[ q\p_q +(q\p_q \ln Z_{h_p}) +{c\over 24} \right]\langle L_{-p}L_p\rangle~,
\eea
where we have assumed $n=-p<0$, the other case leading to the same conclusion. Using our results above, we see that the middle term dominates and implies $\langle L_0 L_{-p}L_{p}\rangle \sim \left({L\over \beta}\right)^5$.  Hence we see that the appearance of an $L_0$ insertion in the third term of (\ref{nnmm_rearrange}) leads to an $L/\beta$ enhancement compared to the first two terms.  The same enhancement arises from the $[L_n,L_{-n}]\sim  n^3$ contribution in the third term.
Therefore,
\bea
\langle L_{-n} L_n  L_{-m} L_m\rangle&\approx& \langle L_{-m}
L_m\rangle \cdot \frac{q^n}{1-q^n}
\left( 2n q\p_q \ln Z+
\frac{c}{12}(n^3-n) \right)\cr
&\approx&\langle L_{-m} L_m \rangle \langle
L_{-n} L_n\rangle,
\eea
or
\be
\langle X_m X_n\rangle \approx \langle X_m \rangle \langle
X_n\rangle(1 + O\left(  { \beta /L} \right))
\ee
for $m\neq n$  in this regime.

Returning to (\ref{offdiag}), using \eqref{nn} and accounting for the two extra powers
of $L/\beta$ that come from replacing the sums by integrals, we have
\be
\delta X^2_\text{off-diag} \sim {1\over L}~,
\ee
for all $w$ at high temperatures.

To compute $\delta X^2_\text{diag}$, we need to evaluate (\ref{nnmm_rearrange})
when $m=n$. In this case, the first and third terms are equal and so
\be
\langle X_n^2\rangle \approx 2\langle X_n\rangle^2.
\ee
This yields the same scaling as for the off-diagonal piece,
\be
\delta X^2_\text{diag} \sim {1\over L}~.
\ee
Altogether, we find that
\bea
\delta \langle \psi_h|T(w)T(0)|\psi_h\rangle=\delta X = \sqrt{ \delta X^2_\text{off-diag}+\delta X^2_\text{diag} }  \sim {1\over \sqrt{L}}~.
\eea
This implies the fluctuation in the correlator
vanishes in the large $L$ thermodynamic limit.

It is straightforward to derive explicit expressions for the fluctuations, analogous to the case of the current correlator. In the limit $L\to \infty$ with $w$ and $\beta$ fixed, substituting $\langle X_n \rangle$ from \eqref{LL}, we have
\begin{align}
\left(\frac{L}{2\pi}\right)^8 \delta X^2_\text{diag}&= \left(\frac{c}{12}\right)^2 \sum_{n>0} \left(\frac{n^3 +\tfrac{L^2}{\beta^2}n }{e^{2\pi \beta/L }-1}\right)^2 \cos^2 \left(\frac{2\pi n
	w}{L}\right)~ \cr
&\approx \left(\frac{c}{12}\right)^2 \left(L \over 2\pi \beta\right)^7 \int_0^\infty \left( \frac{x^3+4\pi^2 x}{e^x-1} \right)^2 \cos^2\left(\frac{w}{\beta}x\right) dx~ \cr
&= \left(L \over  \beta\right)^7 g(w/\beta)~.
\end{align}
The fluctuations in the case with $w/L$ fixed can also be treated as for the current correlator.

Higher-point correlation functions of the stress-tensor  take the
form
\bea \label{tn}
\bra{\psi_h} T(w_1)\dots T(w_n)\ket{\psi_h} = 
(-1)^n\left({2\pi \over L}\right)^{2n} \sum_{\substack{i_1\dots i_n\\\sum i_k =
		0}}  \bra{\psi_h} L_{i_1}\dots L_{i_n}\ket{\psi_h} e^{{2\pi i \over L}\sum_p i_p w_p}.
\eea
It is implicit in the above expression that the $L_0$'s are shifted by $-c/24$.
Equality between (\ref{tn}) in a typical state and its thermal value will follow if the sum is sharply peaked over the ensemble.
We will show in  appendix \ref{appendix_indep} that the fluctuations in (\ref{tn}) are
again small as long as the number of stress tensor insertions is small compared to $L/\beta$.  It  then follows that equality of thermal and typical correlation
functions extends to $n$-point functions of the stress tensor
\bea
\label{independence}
\bra{\psi_h} T(w_1)\dots T(w_n)\ket{\psi_h} &\approx& \langle T(w_1)\dots T(w_n)\rangle_{\beta}~.
\eea

The above arguments hold when the number of stress tensor insertions $n$ is held fixed as $L/\beta \rt \infty$, but can fail if $n$ is allowed to grow in the limit.  This can understood on general grounds as follows.  We write the thermal correlation function of $n$ stress tensors as
\bea
\langle T(w_1) \ldots T(w_n)\rangle = \int_0^\infty \! dE \rho(E) \overline{ \langle E|  T(w_1) \ldots T(w_n)|E\rangle} e^{-\beta E}~,
\eea
where $\overline{ \langle E|  T(w_1) \ldots T(w_n)|E\rangle}$ denotes the average over all states of energy $E$, and $\rho(E)$ is the density of states.  At high temperature, we think of evaluating the integral by locating a saddle point.  Since $\overline{ \langle E|  T(w_1) \ldots T(w_n)|E\rangle}\sim E^n$, if $n$ is held fixed as $\beta \rt 0$ the saddle point location is unaffected by the presence of the stress tensors.   The fact that the same saddle point energy arises independent of the length of the string, provided it is held fixed, is what is responsible for the factorization properties that imply small fluctuations. On the other hand, if $n\sim L/\beta$ (or any more rapid growth) then the saddle point location does depend on the size of the string and  the location of the stress tensors.   Such correlators will therefore be sensitive to the particular microstate, which is not surprising given that in this regime we can arrange the stress tensors uniformly across the system with a spacing less than the thermal wavelength $\lambda \sim \beta$.

\section{Discussion}
\label{discussion}

The main result of this paper confirms a general physical expectation:
correlation functions in typical
high energy states appear thermal.  To reach this conclusion we
needed to be sufficiently careful about what constitutes a typical
state.  A number of past works \cite{Faulkner:2017hll,Basu:2017kzo,Guo:2018pvi,Dymarsky:2018iwx,Maloney:2018yrz,Maloney:2018hdg,Dymarsky:2018lhf,Anous:2019yku} have compared expectation values  in
primary states to those in the thermal ensemble, and in some cases
agreement was found.   As we have seen here, the agreement in these
cases requires working in the large $c$ limit, since at finite $c$
primary states are highly atypical.  This atypicality is responsible
for the mismatch between the expectation values of KdV charges
computed in primary states versus the canonical ensemble.  Typical
states are instead descendants at level $h/c$, and taking this into
account restores the agreement.   We focussed here on correlation
functions of conserved currents and stress tensors, but these remarks
apply generally to correlators computed away
from the large $c$ limit.

Our results have nontrivial implications for the comparison between
CFT and black hole physics. Quantities computed
in a black hole background are inherently coarse-grained and should
therefore be compared with those evaluated in typical states of
the CFT, rather than in primary states. For example, we expect disagreement between correlation
functions in a heavy primary state (as studied e.g.~in
\cite{Fitzpatrick:2014vua,Asplund:2014coa,Fitzpatrick:2015zha,Hijano:2015qja})
and the corresponding Witten diagrams or HRRT surfaces evaluated in the black
hole background beyond leading order in $1/c$.

We have studied the case of the stress tensor in 2D CFT, whose
correlation functions are fixed by conformal
symmetry. One might expect a similar result to hold for generic few-body
operators $O$, namely $\bra{\psi_h} O(w)
O(0)\ket{\psi_h}\approx \langle O(w) O(0)\rangle_{\beta}$, but these
correlation functions depend also on the OPE data of the
theory. However,
conformal symmetry constrains some of the this data
\cite{Kraus:2016nwo,Cardy:2017qhl,Kraus:2017kyl,Das:2017cnv}, which
might lead to approximate equality.  In higher dimensions we lose the power
of Virasoro symmetry and the ability to precisely characterize a
typical state, but the number of descendants still grows exponentially
with the level, and global primaries are more symmetric than
generic operators, so it is plausible that global primaries are atypical in
generic CFTs.

We conclude with a few comments about the connection to the eigenstate thermalization hypothesis (ETH).   The usual statement of ETH is that energy eigenstates of chaotic systems obey \cite{Srednicki:1994,rigol2008thermalization}
\bea\label{ETH}
\langle E_a |O|E_b\rangle =  \langle O\rangle_{\beta_E}\delta_{a,b} + e^{-S(E)/2} f(E_a,E_b)R_{ab} ~,
\eea
where $E$ denotes the average of the nearby energies $E_a$ and $E_b$,
$\langle O\rangle_{\beta_E}$ is the thermal average of the ``few-body"
operator $O$ at the temperature $\beta_E$, $f(E_a,E_b)$ is smooth
function of the energies, and $R_{ab}$ is a random matrix.   Although
the full range of validity of this ansatz remains to be understood, it leads
to physically reasonable behavior regarding the approach to thermal
equilibrium in generic states.   Our results are perfectly compatible
with ETH, and further imply agreement between the vacuum block contribution to CFT quantities
(such as the entanglement entropy) in thermal and
typical states. However, since the only operators $O$ that we study are
conserved currents and the stress tensor, we are not really testing
the core elements of ETH.  For example, the second term in
(\ref{ETH}) is not respected by taking $O$ to be the stress
tensor, since the stress tensor has a strictly vanishing matrix element
between states in different conformal families.

The ETH ansatz ensures that the expectation value of a local operator
averaged over a long time will agree with its thermal value.  In
particular, even if one chooses an initial state for which an expectation
value is far from thermal, the expectation value will simply fluctuate around its thermal value for almost all times, provided the matrix elements of the operator satisfy ETH.     Such time-dependent behavior of course requires the system to be in a non-energy eigenstate (though with a sharply distributed energy), with the time dependence coming from the off-diagonal terms in (\ref{ETH}). In this paper we have restricted attention to energy eigenstates, and although we have considered time dependent correlators this time dependence refers to the relative, as opposed to overall,  location of the operators.  Thus questions regarding thermalization are beyond our present scope, but under investigation.

\section*{Acknowledgements}

We thank Diptarka Das and Mark Srednicki for useful discussions.   P.K. is supported in part by NSF grant PHY-1313986.

\appendix

\section{Current two-point function}
\label{appA}

\subsection{Equivalence of two forms of thermal correlator}
\label{JJ-forms}

Here we establish the equivalence of  (\ref{curra}) and (\ref{currb}).    We start working on (\ref{curra}) by carrying out the sum over over $m$   using
\bea
\sum_m {1\over (2\pi m+a)^2} =   {1 \over 4\sin^2(a/2)}~,
\eea
along with
\bea
E_2(\tau) =1 -6 \sum_{n=1}{1\over \sinh^2 (\pi \beta n/ L)}~,
\eea
the latter following from the identity  $\sum_{n=1}^\infty {nx^{n+1}  \over (1-x^{n+1})^2} =  \sum_{n=1}^\infty {nx^{2n} \over (1-x^n)^2}$.  This gives
\bea\label{corres}
 \langle J(w)J(0)\rangle_{L,\beta} =  {\pi \over \beta L} +{\pi^2\over L^2}  \sum_{n=-\infty}^\infty  {1\over \sinh^2({\pi(\beta n-iw)\over L})} ~.
 \eea

Next, we turn to the mode sum version
\bea
&&\langle J(w)J(0) \rangle_{L,\beta} = \left({2\pi \over L}\right)^2 \sum_n \langle \alpha_n \alpha_{-n} \rangle_{L,\beta}  e^{{2\pi i n w  \over L}}~, \cr
&& =  \left(\tfrac{2\pi}{L}\right)^2 \langle \alpha_0^2\rangle_{L,\beta} + 2\left(\tfrac{2\pi}{L}\right)^2\sum_{m>0} \sinh(\tfrac{\pi \beta m}{ L}) e^{-{\pi \beta m\over L} } \sum_{n=0}^\infty e^{-{2\pi \beta m n\over L}}\left[(n+1) e^{-{2\pi im w\over L}} + ne^{-{2\pi im w\over L} } \right]m~. \cr &&
\eea
We have assumed $0< {\rm Im}(w)<\beta$, so that the sums converge.
Using
\bea
\langle \alpha_0^2 \rangle_{L,\beta} = {\int \! d\alpha_0 e^{\pi i (\tau - \taub) )\alpha_0^2 }  \alpha_0^2\over \int \! d\alpha_0 e^{\pi i (\tau - \taub)\alpha_0^2} }={L\over 4\pi \beta}
\eea
and performing the sum over $n$, we find
\bea
\langle J(w)J(0) \rangle_{L,\beta} = {\pi \over \beta L } +  \left({2\pi \over L}\right)^2  \sum_{m>0}   { \cosh \left( {2\pi m \over L} ({\beta \over 2}-iw) \right) \over \sinh({\pi \beta m\over L} ) } m~.
\eea
Finally, from
\bea
\sum_{m>0}   { \cosh \left({2\pi m \over L} ({\beta \over 2}-iw) \right) \over \sinh\left({\pi \beta m\over L} \right) }  m &= &
\sum_{m>0} \left[ \sum_{n=-\infty}^{-1}  e^{{2\pi m \over L}(iw+n\beta)}m  +  \sum_{n=0}^\infty e^{-{2\pi m \over L}(iw+n\beta)}m\right]\cr& = &
 {1\over 4}  \sum_{n=-\infty}^\infty  {1\over \sinh^2({\pi(\beta n-iw)\over L}) }~,
 \eea
we arrive at (\ref{corres}).

\subsection{Minimal size of thermal correlator}
\label{minimal-size}

We are interested in taking $L\rt \infty$ with $w=L/2$ at fixed $\beta$. This gives the minimal size of the thermal correlator, since periodicity under $w\cong w+L$ implies symmetry around this point.

We proceed by first  performing the over  $n$ in (\ref{curra}), which yields
\bea
\langle J(w)J(0)\rangle_{L,\beta}= -{\pi^2 \over 3L^2 }E_2(\tau) +{\pi \over \beta L}-{\pi^2 \over 3\beta^2}+{2\pi^2 \over \beta^2}\sum_{m=1}^\infty {1\over \sinh^2 ({ L m \over \beta}) }-{\pi^2\over \beta^2} \sum_m {1\over \sinh^2 ({w+ mL \over  \beta}) }~.
\eea
We have the modular transformation
\bea
 E_2(-1/\tau) = \tau^2 E_2(\tau) + {6\tau \over i\pi }~.
 \eea
From this we deduce
\bea
 E_2(\tau) \approx -{L^2 \over \beta^2}+{L \over \pi \beta} +\ldots~, \quad  \beta \rt 0
\eea
where $\ldots$ are exponentially suppressed.  This gives
\bea
 \langle J({L/2})J(0)\rangle_{L,\beta}  \approx  -{\pi\over \beta L} +\ldots~.
 \eea

\section{Technical results}
\label{appB}

\subsection{Higher point functions of the stress tensor}
\label{appendix_indep}
The higher-point functions of the stress tensor in the microstate take the form
\bea\label{higher-point}
\bra{\psi_h} T(w_1)\dots T(w_n)\ket{\psi_h} =
(-1)^n\left({2\pi \over L}\right)^{2n} \sum_{\substack{i_1\dots i_n\\\sum i_k =
		0}}  \bra{\psi_h} L_{i_1}\dots L_{i_n}\ket{\psi_h} e^{{2\pi i \over L}\sum_p i_p w_p},
\eea
analogous to \eqref{TTstate} for the 2-point case. At finite $h$ each $L_0$ should be
replaced with $L_0 - c/24$, but the difference
is subleading in the thermodynamic limit. In order to demonstrate
approximate equality between the microstate and thermal correlators
(\ref{independence}) we must show that this quantity is sharply peaked over
the ensemble of states at fixed $h$. Accordingly, we study the
fluctuations of

\be
Y \equiv \left(\frac{2\pi}{L}\right)^{2n}\sum_{\substack{i_1\dots i_n\\\sum i_k =
		0}}  \bra{\psi_h} L_{i_1}\dots L_{i_n}\ket{\psi_h}e^{{2\pi i \over L}\sum_p i_p w_p}.
\ee
Once again the fluctuations can be split into
off-diagonal and diagonal pieces:
\be
\label{Y}
\delta Y^2 = \langle Y^2\rangle -\langle Y\rangle^2 = \delta
Y^2_\text{off-diag} + \delta Y^2_\text{diag}.
\ee
Here off-diagonal refers to a term of the form
\be
\langle L_{i_1} \dots L_{i_n} L_{j_1} \dots L_{j_n}\rangle~,
\ee
with all of the $i_k$ distinct from all of the $j_k$. As in the main
text, the kinematic factors just go along for the ride.

The diagonal terms are subleading in
the thermodynamic limit, as in section~\ref{statistical}. To see this we make use of a result (proven below) on the expectation values of strings of Virasoro generators. Suppose that $X$ is a string of Virasoro generators of length $\ell$ whose mode
numbers sum to zero, with $s$ the largest number of non-overlapping
substrings within $X$ whose mode numbers sum to zero. If the levels of the generators in $X$ scale like $L/\beta$ and $\ell \ll L/\beta$, then
\be
\label{X}
\langle
X\rangle\sim(L/\beta)^{\ell+s},
\ee
as $(L/\beta)\rightarrow \infty$. These are the levels that are
relevant in the thermodynamic limit, as in the main text. For such
$X$, it also follows that
\bea \label{L0_sub_X}
\langle L_0 X\rangle =  \left[ q\p_q +(q\p_q \ln Z) +{c\over 24} \right]\langle X\rangle \approx  \langle X\rangle\cdot q\p_q \ln Z~.
\eea

First consider the scaling of the connected part of an off-diagonal term,

\be
\left(\frac{2\pi}{L}\right)^{4n}\sum_{\substack{\{i\},\{j\}\\\sum i_k = \sum j_k=
	0}} \langle L_{i_1} \dots L_{i_n} L_{j_1} \dots L_{j_n}\rangle.
\ee
The expectation value scales as $(L/\beta)^{2n+s+s'}$, where $s$
($s'$) is the number of zero substrings in $\{i\}$ ($\{j\}$). We get
additional factors of $(L/\beta)$ when we convert the sums to
integrals: $(L/\beta)^{n-s}$ and $(L/\beta)^{n-s'}$ from the sums over
$\{i\}$ and $\{j\}$ respectively. Accordingly this term scales as
$L^0$,
and one can check that the disconnected piece scales in the same
way. These terms will make an $O(1)$ contribution to $\delta Y^2$
unless they cancel.

Now consider a diagonal term, say with $i_1 = j_1$. The expectation value still
scales as $(L/\beta)^{2n+s+s'}$ but there is one fewer sum since we
have fixed $i_1 = j_1$. This term therefore scales as $L^{-1}$
and vanishes in the thermodynamic limit. Other
diagonal terms will similarly make vanishing contributions to $\delta
Y^2$ in the limit.

Returning to the off-diagonal terms, we see that $\delta Y^2$ will be
$O(1)$ unless
\be
\langle L_{i_1} \dots L_{i_n} L_{j_1} \dots L_{j_n}\rangle \approx
\langle L_{i_1} \dots L_{i_n}\rangle\langle L_{j_1} \dots L_{j_n}\rangle,
\ee
which we now demonstrate. We start from
\be
\langle L_{i_1} \dots L_{i_n}\rangle= \frac{1}{q^{i_1}-1}\sum_{k=2}^n \langle L_{i_2}\dots [{L}_{i_k}, L_{i_1}]\dots L_{i_n}\rangle~,
\ee
where we used the manipulations from
section~\ref{statistical}. Similarly,
\begin{align}
\langle L_{i_1} \dots L_{i_n} L_{j_1} \dots L_{j_n}\rangle = \frac{1}{q^{i_1}-1}&\left[\sum_{k=2}^n \langle L_{i_2}\dots [{L}_{i_k}, L_{i_1}]\dots L_{i_n} \prod_{\ell=1}^m L_{j_\ell}\rangle\right.
\cr &
\left.+\langle \prod_{k=2}^n  L_{i_k} \sum_{\ell=1}^m L_{j_1}\dots [{L}_{j_\ell}, L_{i_1}]\dots L_{j_m} \rangle \right].
\end{align}
The second term has the same length as the first but one fewer zero substring, so by \eqref{X} it has one fewer power of
$(L/\beta)$. The first term therefore gives the leading behavior in the thermodynamic limit:
\bea
\label{ncov}
\langle L_{i_1} \dots L_{i_n} L_{j_1} \dots L_{j_m}\rangle &\approx& \frac{1}{q^{i_1}-1} \sum_{k=2}^n \langle L_{i_2}\dots [{L}_{i_k}, L_{i_1}]\dots L_{i_n} \prod_{\ell=1}^m L_{j_\ell}\rangle.
\eea
This procedure can be iterated on all the $L_{i_k}$ until one is left with only terms of the form
\be
\langle L_{\sum i_k} \prod_{\ell=1}^n L_{j_{\ell}}\rangle \approx \langle \prod_{\ell=1}^n L_{j_{\ell}}\rangle\cdot q\p_q \ln Z ,
\ee
where we made use of \eqref{L0_sub_X}. Thus the expectation value of
the $j$ string factors out:
\bea
\label{factorization}
\langle L_{i_1} \dots L_{i_n} L_{j_1} \dots L_{j_n}\rangle&=&
\frac{1}{q^{i_1}-1}
\sum_{k=2}^n
\langle
L_{i_2}\dots
[{L}_{i_k},
L_{i_1}]\dots
L_{i_n}
\rangle
\langle
L_{j_1}\dots
L_{j_n}
\rangle
(1+O(\beta/L))
\cr
&=&\langle L_{i_1}\dots L_{i_n} \rangle\langle L_{j_1}\dots L_{j_n} \rangle (1+O(\beta/L)).
\eea
We see that the leading term cancels, and the
off-diagonal contribution to $\delta Y^2$ starts at $O(L^{-1})$. This
gives rise to a fluctuation $\sim
\frac{1}{\sqrt{L}}$ at finite size, as for the two-point function.

When the number of insertions scales with $L/\beta$ they can be
arrayed across the entire system with separation smaller than the thermal wavelength, so we
have a very fine-grained probe. In this limit the argument above
breaks down: eq. \eqref{X} no
longer holds and the $q\p_q \langle X\rangle$ term
in \eqref{L0_sub_X} cannot be discarded. Thus the expectation value of the $j$ string does
not factor out, and $Y$ has $O(1)$ fluctuations across the ensemble:
such high-point correlators depend sensitively on the
details of the microstate.

\subsubsection*{Proof of equation (\ref{X})}

Suppose that $X$ is a string of Virasoro generators of length $\ell$ whose mode
numbers sum to zero, with $s$ the largest number of non-overlapping
substrings within $X$ whose mode numbers sum to zero. We will show
that if the levels of the generators in $X$ scale as $L/\beta$ and $\ell \ll L/\beta$, then

\be
\langle
X\rangle\sim(L/\beta)^{\ell+s}
\ee
in the thermodynamic limit.

We proceed by induction; the base case was shown in section~\ref{statistical}. Now, suppose
that the expectation value of an $(\ell,s)$ string scales as
$(L/\beta)^{\ell+s}$ and consider an arbitrary $(\ell+1,s)$ string
\bea
&&\langle L_m L_{a_1}\dots L_{a_\ell}\rangle = \frac{1}{q^m-1}
\sum_{i=1}^{\ell}
\langle L_{a_1}\dots
[{L}_{a_i},L_{m}]
\dots L_{a_\ell}\rangle\cr
&\sim& \sum_{i=1}^{\ell}\left[  (m-a_i)\langle L_{a_1}\dots \hat{L}_{a_i}L_{a_i+m}\dots L_{a_\ell}\rangle -\frac{c}{12}\delta_{m+a_i,0}(m^3-m)\langle L_{a_1}\dots \hat{L}_{a_i}\dots L_{a_\ell}\rangle\right]\cr
&\sim& (L/\beta)^{\ell+s+1},
\eea
provided $m\neq 0$. To obtain the last line we used the
inductive assumption: the first term is the sum of $(\ell,s)$ strings multiplied by one factor of $(m-a_i)\sim (L/\beta)$, while the second term is the sum of $(\ell-1,s-1)$ strings multiplied by $m^3\sim (L/\beta)^3$.

If $m=0$ then we have an $(\ell+1,s+1)$ string
\bea
\langle L_0 L_{a_1}\dots L_{a_\ell}\rangle &\approx& \langle L_{a_1}\dots L_{a_\ell}\rangle\cdot q \p_q\ln Z\cr
&\sim& (L/\beta)^{\ell+s+2}
\eea
where we used (\ref{L0_sub_X}) and the inductive hypothesis. This
proves the claim.

When $\ell \sim L/\beta$ these statements no longer hold: there are
factors $\sim e^{L/\beta}$ relating different orderings of the string,
so there will be some strings contributing to $\delta Y$ for which
\eqref{X}, \eqref{L0_sub_X} and \eqref{factorization} all break down.

\subsection{Ordering independence}

In this subsection we argue that the ordering of operators defining the descendant state does
not affect the expectation value of a string of operators in the thermodynamic limit. This does not affect our results, but leads to an effectively one-to-one correspondence between integer partitions and descendent states for purposes of computing expectation values.

Consider two descendents that differ only in the ordering of two Virasoro generators:
 \be
 \ket{\psi_h} =(L_{\vec{b}})^{\dagger} L_{-i} L_{-j} (L_{\vec{a}})^{\dagger}\ket{h_p},\quad\quad \ket{\psi_h'} =(L_{\vec{b}})^{\dagger} L_{-j} L_{-i} (L_{\vec{a}})^{\dagger}\ket{h_p}.
 \ee
We wish to argue that

\be
\bra{\psi_h} X\ket{\psi_h} \approx \bra{\psi_h'}X\ket{\psi_h'}
\ee
as $\beta/L\rightarrow 0$. First consider $L_{\vec{a}}= L_{\vec{b}}=X=1$:
\bea
\label{norm_h}
\langle{\psi_h}| {\psi_h}\rangle&=&\bra{h_p} L_j [L_i, L_{-i}] L_{-j} \ket{h_p} + \bra{h_p} [L_j, L_{-i}] L_{i} L_{-j} \ket{h_p} \\
&=& f_h(i) f_h(j) +(j+i)\left((j-2i)f_h(j)+(2j-i)f_h(i)+(i-j)f_h(i-j)\right), \nonumber  
\eea
where
\be
f_h(n) = 2nh_p+\frac{c}{12}(n^3-n).
\ee
When $n\sim L/\beta$ and $h\sim (L/\beta)^2$, $f_h(n)\sim (L/\beta)^3$ and so the first term in (\ref{norm_h}) gives the leading thermodynamic behavior $\sim (L/\beta)^6$. On the other hand,
\bea
   \label{reordering2}
   \langle{\psi_h} | {\psi_h}\rangle - \langle{\psi'_h}|  {\psi'_h}\rangle &=&
                             \bra{h_p} L_i L_j
                           [L_{-i}, L_{-j}]
                          \ket{h_p}
                             +\bra{h_p} [L_j, L_i]
                             L_{-i} L_{-j}
                           \ket{h_p}\cr
&=&  (j-i)  \left( (2j+i) f_h(i) + (2i+j) f_h(j)\right)
                             \eea
which scales as $(L/\beta)^5$. This is the key point: terms that arise in the difference $\bra{\psi_h} X \ket{\psi_h} - \bra{\psi_h'} X \ket{\psi_h'}$ have one factor of $f_h$ and two of the levels in place of $f_h^2$ in $\bra{\psi_h} X \ket{\psi_h}$, so the difference is subleading to the expectation values themselves in the thermodynamic limit.

If we now let $L_{\vec{a}}, L_{\vec{b}}$ and $X$ be arbitrary the above reasoning still applies. The objects to compare are
\be
   \label{genreoder}
                            \bra{h_p} L_{\vec{b}} L_j L_i
L_{\vec{a}} X  (L_{\vec{a}})^{\dagger} L_{-i} L_{-j}
L_{\vec{b}})^{\dagger}\ket{h_p}\quad\text{vs.}\quad
\bra{h_p} L_{\vec{b}} L_j L_i
L_{\vec{a}} X (L_{\vec{a}})^{\dagger} [L_{-i}, L_{-j}]
(L_{\vec{b}})^{\dagger} \ket{h_p},
\ee
which can be computed by commuting $L_{-i}$ and $L_{-j}$ (or $[L_{-i},L_{-j}]$) all the way to the left. In the first a term with two factors of $f_h$ is generated, while the second has at most one factor of $f_h$ and two
factors of the levels. The remainder of the computation is the same in both cases, so the difference is subleading in the limit.

\bibliographystyle{bibstyle2017}
\bibliography{collection}

\end{document}